\documentclass{aastex}
\usepackage{emulateapj5}
\usepackage{onecolfloat5}

\input epsf
\def\plotone#1{\centering \leavevmode
\epsfxsize= 1.0\columnwidth \epsfbox{#1}}

\def\apjl{Astrophys. J. Lett.}
\def\mnras{Mon.Not.Roy.As.Soc.}
\def\araa{Annu. Rev. Astron. Astrophys.}
\def\aj{Astron. J.}

\makeatletter

\newcommand{\tableskip}{\\[-6pt]}

\def\be{\begin{equation}}
\def\ee{\end{equation}}
\def\bea{\begin{eqnarray}}
\def\eea{\end{eqnarray}}

\def\cmm2{{\,\rm cm^{-2}}}
\def\cm2{{\,{\rm cm}^2}}
\def\cmm3{{\,{\rm cm}^{-3}}}
\def\gcmm3{{\,{\rm g\,cm^{-3}}}}

\def\HO{{100h\,{\rm km\,sec^{-1}\,Mpc^{-1}}}}

\def\fun#1#2{\lower3.6pt\vbox{\baselineskip0pt\lineskip.9pt
  \ialign{$\mathsurround=0pt#1\hfil##\hfil$\crcr#2\crcr\sim\crcr}}}
\def\C{{\cal C}}

\def\muK{\mu {\rm K}}

\def\w{\omega}

\def\p3m{P$^3$M}

\def\la{\mathrel{\mathpalette\fun <}}
\def\ga{\mathrel{\mathpalette\fun >}}
\def\fun#1#2{\lower3.6pt\vbox{\baselineskip0pt\lineskip.9pt
  \ialign{$\mathsurround=0pt#1\hfil##\hfil$\crcr#2\crcr\sim\crcr}}}

\font\BF=cmmib10

\def\v{{\hbox{\BF v}}}

\begin{document}
\bibliographystyle{apj}
\twocolumn[
\submitted{Submitted to ApJ}
\title{Rapid Calculation of Theoretical CMB Angular Power Spectra}
\author{Manoj\ Kaplinghat, Lloyd\ Knox and Constantinos\ Skordis}
\affil{Department of Physics, One Shields Avenue\\
University of California, Davis, California 95616, USA}

\begin{abstract}
We have developed a fast method for predicting the angular power
spectrum, $C_l$, of the cosmic microwave background given
cosmological parameters and a primordial power spectrum of
perturbations.  After pre--computing the radiation temperature
and gravitational potential transfer functions
over a small sub--space of the total model
parameter space, the rest of the model space (six or more
cosmological parameters and arbitrarily many primordial power
spectrum parameters) is reached via rapid analytic and
semi--analytic approximations which are highly accurate on all
angular scales for which linear perturbation theory applies.  A
single power spectrum can be calculated in $\sim 1$ second on
a desktop computer. We discuss applications to cosmological parameter
estimation.
\end{abstract}
 ]

\section{Introduction}

The anisotropy of the Cosmic Microwave Background (CMB) is proving
to be a powerful cosmological probe.  Measurements of its angular
power spectrum can be used to tell us about the baryon density,
dark energy density, the nature of the dark matter, the age of the
universe and the primordial spectrum of perturbations generated in
the inflationary era
\citep{pryke01,netterfield01,lee01,wang01,knox01}.

A persistent challenge to the analysis is the large number of
model angular power spectra ($C_l$) that must be calculated in
order to understand the constraints the data place on parameter
spaces with seven to ten or even higher dimensions.  Here we
present a fast, yet accurate, method for computing the $C_l$ for a
given model.

These model angular power spectra are the expectation values of
the variance of spherical harmonic coefficients, $a_{lm}$, where
$\langle a_{lm}a^*_{l'm'} \rangle = C_l \delta_{ll'}
\delta_{mm'}$. The $C_l$ depend on 
the density of dark matter, the fraction of this
which is hot dark matter, the density of baryonic matter,
the redshift of reionization of the intergalactic medium, the dark
energy density, the dark energy pressure and the mean spatial
curvature. These cosmological parameters influence the evolution
of perturbations in the photon temperature.  The $C_l$'s also
depend on the statistical properties of the initial perturbations,
possibly produced in an epoch of inflation. These initial conditions
are described with the primordial gravitational potential power
spectrum, $P(k)$.

The $C_l$ can be calculated highly
accurately because of the applicability of linear perturbation
theory. Indeed, this is one of the reasons the CMB is such a
powerful cosmological probe. One need only solve the linearized
Einstein and relevant Boltzmann equations, which can be cast as a
set of coupled ordinary differential equations.  Early codes (e.g.
\citet{bond84}) directly solved the whole hierarchy, up to some
limiting multipole moment, of the photon temperature perturbation
and could take tens of hours to calculate $C_l$ for a single model.

\citet{hu95} introduced a semi--analytic approach 
which was much faster than the ``whole
hierarchy'' Boltzmann codes of the day and with an accuracy around
10\%.  That accuracy could be improved, but only at the expense of
much slower performance.

The line--of--sight integration method for solving the linearized
Einstein and Boltzmann equations \citep{seljak96} greatly reduced
the time required for calculation of accurate theoretical power
spectra by bypassing the need to solve the
whole hierarchy. Publicly available codes based on this method,
mostly CMBfast \citep{seljak96}, have been the workhorses of all
parameter--determination efforts to date. Despite its great speed,
these analysis efforts have typically required months of running
CMBfast.

\citet{tegmark00} introduced a high--$\ell$ / low--$\ell$ split
in the calculation of $C_l$ to exploit analytic approximations
valid at high $\ell$ and insensitivity to certain parameters at 
low $\ell$.  With this split they were able to calculate a 7--dimensional
grid of $C_l$'s with many fewer calls of CMBfast than would have
been required for a brute--force calculation.  By analytically correcting
these for reionization effects and scaling the tensor and scalar power spectra
with separate amplitudes they covered a 10--dimensional parameter space.

We use a similar high--$\ell$ / low--$\ell$ split to exploit the
same analytic approximations as \citet{tegmark00}.  
Our methods though have several advantages including
reduced pre--compute time ($\sim 1$ month reduced to $\sim 40$ hours), 
and ability to handle large numbers of primordial power spectrum parameters.
We have performed extensive accuracy tests showing
that our errors are smaller than cosmic variance errors for $l < 1000$.  
We achieve these advantages by storing the Fourier-- and Legendre--transformed
temperature perturbation transfer function (rather than $C_l$), 
using more efficient choices for grid parameters, and further use
of (highly accurate) analytic approximations.  We also have
an option where all the low--$\ell$ effects are calculated with
semi--analytic approximations, greatly reducing pre--compute times
and storage requirements even further, as well
as allowing for greater ease in incorporating new physical effects.
Our software package is called
the Davis Anisotropy Shortcut (DASh)\footnote{DASh can be downloaded from \\
http://www.physics.ucdavis.edu/Cosmology/dash/}.

DASh incorporates many approximations that have been presented 
elsewhere in the literature, and also some new ones.
We present an approximate scaling for the polarization low-$\ell$
reionization feature, an improved approximation for the reionization
damping factor and a generalization of the angular--diameter 
distance scaling which makes it accurate for all angular scales 
even in the presence of non--zero curvature.  We also present
improved semi-analytical approximations for the calculation of
low-$\ell$ temperature spectra. These improvements enable a
semi-analytical calculation of low-$\ell$ spectra to an accuracy 
better than 2$\%$ on average over a wide range of cosmological
parameters (including curvature).

Although recent work has shown the exploration of these large
model spaces to be possible without DASh, our method greatly
reduces the required computer resources.  As such it will allow for
extension to more parameters including those needed to describe
isocurvature components, or the dark energy pressure. A particulary
straightforward extension would be to the number of parameters used to
describe $P(k)$, beyond the usual two needed for the power--law
description.  A faster method also makes it possible to redo
calculations to check for sources of systematic error.

A preliminary version of DASh has already been used for parameter
estimation from CMB data \citep{knox01}.  There we combined DASh
with the Monte Carlo Markov Chain (MCMC) approach to Bayesian
inference described in \citet{christensen01}.  The MCMC approach
requires many fewer likelihood evaluations than a direct
grid--based approach even for applications with only a handful of
parameters, and generally becomes even more advantageous as the
dimensionality increases further \citep{gilks96}.  Others have
used MCMC for cosmological problems \citep[e.g.][]{verde02} and
we expect the technique to become widely used in cosmology.

CMB anisotropies are conveniently broken up into two different types:
those which are simply projections of features on (or near) the
last--scattering surface (early anisotropy) and those that are
generated much more recently (late anisotropy).  After reviewing some
notation in Section 2 we discuss our computation of early anisotropies
in Section 3.  In Section 4 we describe the two different ways we
compute late--time effects due to gravitational potential decay and
reionization of the inter--galactic medium. In Section 5 we quantify
the level of accuracy by comparing 6,823 models as calculated with
DASh to those calculated with CMBfast.  In section 6 we describe our
calculation of polarization power spectra and the contribution from
tensor perturbations.  In Section 7 we consider extensions, for example 
to including lensing effects, and finally in Section 7 we conclude.

\section{Notation}

Before discussing the method we quickly review some notation.  The
temperature observed in direction ${\hat \gamma}$ observed from
any point in space, ${\bf x}$, can be written as \be T({\bf
x},\hat\gamma) = \bar T + \Delta({\bf x},\hat\gamma).\ee For
anisotropy sourced by scalar metric perturbations the
Fourier--transformed temperature perturbation is azimuthally
symmetric and can be expanded in Legendre polynomials as \be
\Delta({\bf
k},\hat\gamma)=\sum_l\left(2l+1\right)(-i)^l\Delta_l({\bf k})P_l(\mu)
\ee where $\mu = \hat k \cdot \hat \gamma$.
The multipole moments of the Fourier--transformed temperature
perturbation can be written as
$\Delta_l({\bf k})=\Psi_i({\bf k})\Delta_l(k)$ where
${\bf k} = k \hat k$ and  $\Psi_i({\bf k})$ is the perturbation in the
gravitational potential \citep{ma95} at some very early time when all
relevant perturbation wavelengths are larger than the horizon.  Note
that when we write $\Delta_l(k)$ with a scalar rather than vector
argument (as we do throughout),  we are using it as a transfer
function.

If we solve for $\Delta_l(k)$ assuming adiabatic initial conditions
with $\Psi_i({\bf k})=1$,
then if we assume the perturbations are statistically isotropic and
homogeneous we can calculate $C_l$ for any arbitrary initial
potential power spectrum $P(k)$ as
\be
\label{eqn:dlk2cl} C_l = (4\pi)^2\int_0^\infty dk k^2 \Delta_l^2(k)P(k)
\ee
where $C_l$ is defined by
\be\langle a_{lm}({\bf x}) a^*_{l'm'}({\bf x}) \rangle =
C_l \delta_{ll'}\delta_{mm'} \ee and
\be a_{lm}({\bf x}) = \int d\hat\gamma Y_{lm}(\hat \gamma)
\Delta({\bf x},\hat\gamma). \ee

We often express densities in units of the critical density for
$h=1$ where $H_0 = \HO$ and the critical density is $\rho_c \equiv
3H_0^2/(8\pi G)$.  Following convention, we refer to densities in
these units with the symbol $\w$.  The baryon density is $\w_b$,
the dark matter density is $\w_d$, the matter density is $\w_m =
\w_b+\w_d$, and the dark energy density is $\w_x$.  Note that
$\w_i = \Omega_i h^2$.  These symbols all refer to present day
densities.  We define a curvature ``density'' as $\w_K \equiv
\Omega_K h^2=(1-\Omega_{\rm tot})h^2$ where $K=-1,+1,0$
corresponds to an open, closed or flat universe respectively. With
this definition the Friedmann equation at the present time becomes
$h^2 = \sum_i \w_i$.  We assume that a fraction, $f_h$, of the
dark  matter is hot and that the rest is cold.  We further assume
that the dark energy is a cosmological constant, though we discuss an
extension of DASh to $w_x \equiv p_x / \rho_x \ne -1$ models.

\section{Early Anisotropies}

The dynamical processes at early times (e.g., acoustic
oscillations of the baryon--photon fluid, Hydrogen and Helium
recombination rates and Silk--damping) are governed only by
$\w_b$, $\w_m$ and $f_h$. Photon density matters as well, but this
is well--determined from the FIRAS measurement of the CMB
temperature as $T = (2.728 \pm 0.004)$~K (95\% confidence)
\citep{fixsen96}. Dark energy parameters and the curvature radius
are irrelevant since the dark energy density at early times was
negligible (in most models, certainly for a cosmological constant)
and the curvature radius at last--scattering was much larger than
the horizon at that time.

The small number of parameters which are necessary for fixing the
statistical properties of the CMB at early times and on small
scales led \citet{tegmark00} to create a high--$\ell$ grid of
angular power spectra with grid parameters, $n_S$, $\w_b$, $\w_d$
and $f_h$.  Although $\Omega_\Lambda$ and $\Omega_K$ do affect the
projection of comoving length scales into angular scales, they do
so in a particularly simple manner. With the grid constructed at
fiducial values of $\Omega_K = \Omega_K^*$, $\Omega_\Lambda =
\Omega_\Lambda^*$, they obtain $C_l$ for non--fiducial values of
the curvature via \citep{wilson82}:
\be \label{eqn:project}
\C_l(\Omega_K,\Omega_\Lambda,\w_b,\w_m,f_h) = \C_{\tilde
l}(\Omega_K^*,\Omega_\Lambda^*,\w_b,\w_m,f_h)
\ee
where $\C_l \equiv l(l+1)C_l/(2\pi)$,
\be
\tilde l/ l = D_A^{z_{\rm
peak}}(\Omega_K^*,\Omega_\Lambda^*) / D_A^{z_{\rm
peak}}(\Omega_K,\Omega_\Lambda)
\ee
and $D_A^z$ is the angular diameter distance to $z$ and $z_{\rm peak}$
is the redshift where the visibility function peaks. In section 5 we derive
Eq.~\ref{eqn:project} and also a version which does not rely on any
small--angle approximation, as this one does.

Early anisotropy effects for DASh are also calculated via direct
numerical solution of the linearized Einstein and Boltzmann
equations over a grid of parameters.  The key difference is that
DASh stores the Fourier and Legendre--transformed photon
temperature perturbation, $\Delta_l(k)$, instead of $C_l$. Because
of this difference, our grid only needs to contain cosmological
parameters, and not the primordial power spectrum parameters. The
dimensionality of the grid is reduced (and with it the storage
requirements) and flexibility is increased since we are no longer
restricted to power--law descriptions of the primordial power
spectrum.  A typical use of DASh will first take tens of hours of
computing the $\Delta_l(k)$ grid by a call of
CMBfast \citep{seljak96} for each grid point. Only after the entire
grid is computed (we say ``pre--computed'') can DASh produce
angular power spectra in $\sim 1$ second, as advertised.
Specifically, the grid is over parameters $\omega_b$, $\omega_m$
and $f_h$ at fixed values of $\Omega_K \equiv 1-\Omega_{\rm tot}=
\Omega_K^*$, $\Omega_\Lambda = \Omega_\Lambda^*$ and $\tau=0$. For
reasons of algorithmic simplicity the current implementation
requires the number of grid points for each grid parameter to be a
power of 2. From this grid, we get $\C_l$ for any $\omega_b$,
$\omega_m$, $f_h$ and the primordial power spectrum $P(k)$ by performing
multi--linear interpolation on the grid of $\Delta_l(k)$ and then
the integral in Eq.~\ref{eqn:dlk2cl}.  DASh can then get any
$\C_l$, accurate for $l \ga 100$, in the model space of
\{$\omega_b$, $\omega_d$, $\Omega_\Lambda$, $\Omega_K$, $P(k)$\}
via Eq.~\ref{eqn:project}.

Grid boundary and finite grid--spacing effects can be minimized by
an intelligent choice of the parameters.  For example, instead of
gridding uniformly in $\w_m$ we grid uniformly in $\ln{\w_m}$
which makes the interpolation error more uniform over the range of
$\w_m$ values. The uniformity of errors is desirable since if one
holds the number of grid points fixed a parameterization that has
more uniform errors has a smaller largest error.  We discuss variable
choice more in the next section.

\section{Late--time and Geometric Effects}

Although we can use a low--dimensional parameterization of the
early anisotropy, the late anisotropy is sensitive to additional
effects and more cosmological parameters. The additional effects
are due to geometry, the decay of the gravitational potential
which occurs in the curvature or dark--energy dominated era, and
Thomson scattering off of the free electrons in the re--ionized
intergalactic medium.

We take two approaches to including these additional effects. One
approach requires calculation of a second grid of $\Delta_l(k)$ (the
``low--$\ell$ grid'' which has more dimensions than the high--$\ell$
$\Delta_l(k)$ grid already described.  The other relies solely on
semi--analytic calculation for the late--time effects.  The first
we will refer to as gDASh and the second as sDASh.
The sDASh is not completely grid--independent; it relies on the high--$\ell$
$\Delta_l(k)$ grid as an accurate description of the photon
perturbations at early times and sub--curvature scales. 
Below we first describe the
semi--analytic calculation of the various effects and then the
additional grid.

We model the radiation temperature transfer function
as being modified by one additive factor and one multiplicative factor:
\be
\Delta_l(k) = \Delta_l^{\rm ISW}(k) + R_l(\tau) \Delta_l^{\rm early}(k).
\ee
where $\Delta_l^{\rm early}(k)$ is interpolated from the previously
described grid, $\Delta_l^{\rm ISW}(k)$ is the late--time contribution
from the ``Integrated Sachs--Wolfe'' effect explained below and $R_l(\tau)$
is the reionization damping factor for late--time optical depth to
Thomson scattering, $\tau$.
The resulting power spectrum can thus be written as
\be
\label{eqn:master}
C_l = C_l^{\rm ISW} + R_l^2(\tau)C_l^{\rm early} + R_l(\tau)
C_l^{\rm ISW-early}
\ee
In the following subsections we describe how we calculate $C_l^{\rm early}$ (which gets geometric corrections), $C_l^{\rm ISW}$, the $C_l^{\rm ISW-early}$
cross term and $R_l(\tau)$.

\subsection{Geometry}

Although the curvature scale is larger than the horizon at last
scattering, curvature does have effects on the early evolution of
super--horizon size modes, which are unobservable at the time of
last--scattering, but which have observational consequences now.
That this is the case should not be surprising since it is
impossible to map, without deformation, a space of zero mean
curvature onto one with non-zero mean curvature.  One can see this
formally as a result of the fact that the eigenfunctions of the
Laplacian are different in spaces of different curvature. We will always
denote the eigenvalues of the Laplacian as $k$. One can further define a
``wavenumber'' in curved space as $\beta^2=k^2+K/r_{\rm curv}^2$, where
$r_{\rm curv}=H_0^{-1}/\sqrt{|\Omega_K|}$ is the curvature radius.
In the closed case, the spectrum of eigenvalues is
discrete and $\beta r_{\rm curv}$ takes on only integer values. Further,
$\beta r_{\rm curv} =1,2$ are pure gauge modes \citep{bardeen80}. The
eigenfunctions of the Laplacian in curved space are the so--called
hyperspherical Bessel functions. We will follow the definition and
notation of \citet{abbott86} for the hyperspherical Bessel function and
denote them as $\Phi_\beta^l(\chi)$. At small distances and short
wavelengths, $\Phi_\beta^l(\chi) = j_\ell(k\chi)$ and
$k=\beta$. For more details on perturbation theory and CMB
anisotropies in non-flat backgrounds we refer the reader to
\citet{kamionkowski94, white96}.

We take as our starting point for this calculation the
$\Delta_l(k)$ already stored in the high--$\ell$ grid with
fiducial parameter values $\Omega_\Lambda = \Omega_\Lambda^*$ and
$\Omega_K = 0$. Recall that this grid is for the temperature
perturbation {\it today} and not on the last--scattering surface.
Calculating $C_l$ from this grid for arbitrary $\Omega_\Lambda$
and $\Omega_K$ requires two steps. We must first correct for the
effects of curvature at last--scattering, and then correct for how
the projection from last--scattering to today has changed. As
mentioned, curvature introduces a cutoff scale, $k_{\rm curv}$, in
the spectrum of Laplacian eigenvalues such that
$k_{\rm curv}  = {1/r_{\rm curv}, 0, \sqrt{8}/r_{\rm curv}}$ for
${\rm K}={-1, 0, 1}$.
Our correction for the effect of curvature at the epoch of
last--scattering is to introduce a cutoff in the integral over
$k$; i.e., we define
\be \label{eqn:dlk2clf}
C^g_l = (4\pi)^2\int_{k_{\rm curv}}^\infty
dk k^2 \Delta^g_l(k)^2P(k),
\ee
where the $g$ superscript stands for ``grid'' and implies that the
quantity in question has been obtained from the grid by interpolation.
We have found that this simple approximation works very well.
Note that $P(k)$ here is the power spectrum for the flat model. We
then use $C^g_l$ to calculate an intermediate angular correlation
function $C^g(\theta)$ which then needs to be stretched to the
correct angular diameter distance. For monopole (isotropic)
sources emitting from a thin shell the shift is particularly
simple \citep{wilson82}: \bea \label{eqn:projectlarge}
C(\theta) & = & C^g(\theta');\nonumber\\
\xi & \equiv & 2\,D_A^{z_{\rm peak}}(0,0)\sin(\theta'/2),\nonumber\\
\sin(\theta/2) & = & \sinh_K(\xi/2)/
D_A^{z_{\rm peak}}(\Omega_K,\Omega_\Lambda).\label{eqn:shift-theta}
\eea
The function $\sinh_K(x)$ is defined as
$\sin(x)$, $x$, $\sinh(x)$ for ${\rm K}=1,0,-1$ respectively.
Legendre--transforming the shifted $C(\theta)$ back to $\ell$--space
then gives us the $C_l^{\rm early}$ of Eq.~\ref{eqn:master}.

The transformation of the correlation function is only exact for
{\it monopole} sources on a thin shell at fixed redshift.  The dipole
source due to the peculiar velocities of the photon--baryon fluid,
the thickness of the last--scattering surface, and late--time
effects all violate these restrictions.  We discuss the resulting
(very small) errors in section 5.

At small angles ($\theta << 1$), Eq.~\ref{eqn:shift-theta} can be
cast in terms of a shift in $\ell$, as given by
Eq.~\ref{eqn:project} (see section 5 for a derivation).
Thus at small angular scales ($\ell \ga 10$), a simple shift in $\ell$
is sufficient and one need not Legendre transform. 

\subsection{Gravitational Potential Decay}

We now
show how to calculate the $C_l^{\rm ISW}$ term and the
$C_l^{\rm ISW-early}$ cross term of Eq.~\ref{eqn:master}.  These result from
the late--time generation of anisotropy due to gravitational
potential (metric perturbation) decay.  In linear
perturbation theory gravitational potentials are independent of time
when the Universe is completely flat and matter--dominated, but decay in
the presence of curvature and/or dark energy.  As CMB photons pass
through the evolving potentials, new (secondary) anisotropy is created
via what is called the Integrated Sachs-Wolfe (ISW) effect \citep{hu95}.

We calculate the ISW effect by evaluating the line--of--sight integral:
\be
\label{eqn:deltalk-isw} \Delta^{\rm ISW}_l(k)= \int_{\eta_{\rm
late}}^{\eta_0}d\eta \Phi_\beta^l(\chi) S^{\rm ISW}(k,\eta),
\ee
where $\eta_0$ is the conformal time today and
$\chi=\eta_0-\eta$. $\eta_{\rm late}$ is some late time
prior to the onset of curvature domination or
dark energy domination, which ever is earlier.

The ISW source term is:
\be\label{eqn:source-isw} S^{\rm ISW}(k,\eta)=2
e^{-\tau(\eta)} \dot{\Psi},
\ee
where the optical depth to
Thomson scattering is given by
\be
\tau(\eta) \equiv \int_\eta^{\eta_0} \dot \tau d\eta \ \ ; \ \
\dot{\tau}=\bar n_e \sigma_T a
\ee
where $\bar n_e$ is the mean number density of free electrons and
$\sigma_T$ is the Thomson cross section. The visibility function is
$g=\dot{\tau}\exp(-\tau(\eta))$.
The gravitational potential $\Psi$ is defined in \citet{ma95}.

To calculate $\Psi$, when we make the grid of flat models we store
not only $\Delta_l(k)$ but also $\Psi^g(k,z=100)$.  This we use
as a transfer function, and then numerically solve for the growth
factor, $D(z)$ so that
\be
\Psi(k,z)= {D(z)\over D(100)} \Psi^g(k,100)
\ee
This factorization
is possible because the evolution of non-relativistic
matter perturbations is independent
of $k$ for modes inside or outside the horizon when the clustered
components are pressureless \citep{heath77}. For the growth function
$D(z)$, we use the approximation given by \citet{carroll92}.

For values of $\beta$ smaller than some
($l$ dependent) multiple of $l/\eta_0$, we evaluate the intergral in
Eq.~\ref{eqn:deltalk-isw} explicitly.
For other values we use a generalization of the
weak--coupling approximation of \citet{hu96} which works best for large
values of $\ell$ and $\beta$.
Since $\Phi_\beta^l(\chi)$ is a rapidly varying quantity, one can take the
source term out of the integral and evaluate it at the conformal time
where $\Phi_\beta^l(\chi)$ attains its maximum ($\chi_{\rm max} =
\eta_0-\eta_{\rm max}$). This allows Eq.~\ref{eqn:deltalk-isw} to be
written as:
\be\label{eqn:weak-coupling}
\Delta^{\rm ISW}_l(k)\simeq S^{\rm ISW}(k,\eta_{\rm max})\int_0^{\eta_0}d\chi \Phi_\beta^l(\chi) ,
\ee
One is then left with the integral over the hyperspherical Bessel
function whose solution can be written as a recurrence relation; we only
need the values of the integral for $l=0,1$. Denoting the integral in
Eq.~\ref{eqn:weak-coupling} by $I^l_\beta$, the following recurrence
relation can be derived:
\be\label{eqn:recurrence}
I^l_\beta=-\frac{2l-1}{l}\frac{\Phi^l_\beta(\eta_0)}{\sqrt{\beta^2-Kl^2}}
+\frac{l-1}{l}\frac{\sqrt{\beta^2-K(l-1)^2}}{\sqrt{\beta^2-Kl^2}}I^{l-2}_\beta
\ee
For open and flat models analytical solutions can be written by taking
$\eta_0$ to infinity; for closed models, we numerically evaluate
$I_\beta^l$ using Eq.~\ref{eqn:recurrence}.

Now we turn to the $C_l^{\rm ISW-early}$ cross term.  The largest
correlation with late ISW comes from the Sachs--Wolfe (SW) effect.  We
currently neglect contributions from the primary Doppler and early ISW
effects, though including them would improve the accuracy.  The
Sachs--Wolfe radiation temperature transfer function is given by
$\Delta^{\rm SW}_l(k)=[\Theta_0+\Psi]^g(\eta_{\rm lss})
\Phi_\beta^l(\eta-\eta_{\rm lss})$, where ``lss'' stands for Last
Scattering Surface taken to be at $z=1100$ and
$[\Theta_0+\Psi]^g(\eta_{\rm lss})$ is the effective photon
temperature for the corresponding model interpolated from the
high--$\ell$ grid. With $\Delta_l^{\rm early} \simeq \Delta_l^{\rm SW}$
thus calculated, we then get
\be
C_l^{\rm ISW-early} = (4\pi)^2\int_{\beta_{\rm K}}^\infty \beta^2 d\beta
\Delta_l^{\rm ISW}\Delta_l^{\rm early} P_{\rm K}(\beta) ,
\ee
where $P_{\rm K}(\beta)$ is the curved space initial potential power
spectrum \citep{zaldarriaga98}, and $\beta_{\rm K}={0,0,2/r_{\rm curv}}$
for ${\rm K}={-1, 0, 1}$.

\subsection{Reionization}\label{sec:reion}

Thomson scattering smears out our view of the last--scattering
surface, and therefore damps the early anisotropy.  This damping
is described by the reionization damping factor, $R_l(\tau)$.
Our first step to calculating $R_l(\tau)$ is to extract it numerically
from models in a grid (pre--computed with multiple calls to CMBfast)
over $\w_b$, $\w_m$, $f_h$ and $\tau$ with
$\Omega_K = 0$ and $\Omega_\Lambda=0$.  These models have no ISW
effect so we simply set
\be
R_l^2(\w_b,\w_m,f_h,\tau) =
C_l(\w_b,\w_m,f_h,\tau)/C_l(\w_b,\w_m,f_h,0). \nonumber
\ee
As we will see later, at high $\ell$ $R^2_l = \exp(-2\tau)$, so
we actually extract $F_l$ instead where
$R^2_l = F_l (1-\exp(-2\tau))+\exp(-2\tau)$.
The $F_l$ are stored as a function of $\ell/(\ell_r+1)$ and we
interpolate between the stored values to obtain the $F_l$ for a target
model with $\Omega_K = 0$, $\Omega_\Lambda=0$. The reionization
multipole is defined by $\ell_r = D_A^{z(\eta_r)}/\eta_r$, where
$\eta_r$ is the visibility function weighted conformal time
\citep{hu97}. As pointed out by \citet{hu97},
once we have $R_l$ as a function of $\ell/(\ell_r+1)$, it is not changed
significantly by curvature or $\Lambda$ or any other late time effect
that happens {\em after} reionization. The $R_l$ for arbitrary
model parameters are thus given by
$R_l(\Omega_K,\Omega_\Lambda) = R_{\tilde l}(0,0)$ where
$\tilde l = l (l_r(\Omega_K,\Omega_\Lambda)+1)/(l_r(0,0)+1)$.

The reionization damping term $R_l^2(\tau)$ could also be obtained
semi--analytically.  Note that the source
terms for the early anisotropy (not shown in Eq.~\ref{eqn:source-ri})
all get suppressed by $e^{-\tau}$, {\em independent} of $\ell$.
The $\ell$--dependence of the damping factor comes entirely from the
non--ISW late--time creation of anisotropy at low $\ell$ via a
source term
\bea\label{eqn:source-ri}
S^{\rm
RI}(k,\eta)&=&g(\Theta_0+\Psi)\nonumber\\ &+&\frac{2b_k}{3\Omega_m}
\frac{d}{d\eta} \left[g\left(\dot{a}\Psi+a\dot{\Psi}\right)\right]
\eea
where $a$ is the scale factor and $b_k\equiv (k^2-3K/r_{\rm curv}^2)/k^2$.
Thus
\be
R_l^2(\tau)=\frac{C_l^{\rm RI}(\tau)+e^{-2\tau}C^f_l(0)}{C^f_l(0)},
\ee
where $C_l^{\rm RI}$ is calculated using $S=S^{\rm RI}$ for
$z<z_{\rm ri}$ and $S=0$ for $z>z_{\rm ri}$.

In writing $S^{\rm RI}$ we have neglected the Doppler source term
proportional to the difference in baryon and photon fluid velocities.
For reasonable values of the baryon density, this late--time Doppler
effect only starts to become important for $z_{\rm ri} \ga 25$
\citep{hu96}.

We treat reionzation as if it instantaneously occurred at
$z =z_{\rm ri}$. We use $z_{\rm ri}$ instead of $\tau$, as the input to
DASh, since $z_{\rm ri}$ is more directly related to observational 
constraints \citep{becker01,fan01}.  When we sample parameter
space in order to characterize the accuracy of DASh, as described
in Section 5, we always keep $z_{\rm ri} \le 10$. For a review of
theoretical work on reionization see \citet{loeb01}. 

%

\subsection{A low--$\ell$ grid}

We have also implemented in DASh a numerical calculation of the
gravitational potential decay and geometrical effects with the
pre--computation of a low $\ell$ $\Delta_l^2(k)$ grid.  This
second--grid approach, called gDASh, has the advantages of speed and tunable
accuracy over sDASh.

The low--$\ell$ grid, due to its incorporation of the late--time
effects, necessarily has more dimensions than the high--$\ell$ grid.
We have chosen these extra variables to be
$\Omega_\Lambda/\Omega_m$, since this combination controls the ISW
effect and $\w_K$ since this sets the curvature radius.
The low--$\ell$ grid is {\em less} sensitive to $f_h$, $\w_b$ and
$\w_m$ than is the case for the high--$\ell$ grid so we can grid
more coarsely in these \citep{tegmark00}.

Although $\w_m$ has little effect on late--time
generation of anisotropy,  this parameter directly
controls the amount of {\it early} ISW effect.  Since the early
ISW effect is not projected to us from the last--scattering
surface, the angular scaling assumed for use of the early grid
will introduce some errors.  Fortunately, these errors are
negligible and the late grid can indeed be fairly coarse in
$\w_m$.

We join the results of the low--$\ell$ 
and high--$\ell$ calculations by simply using the
low--$\ell$ calculation up to a limiting value $l_{\rm late}$.  Our
algorithm for choosing $l_{\rm late}$ is derived from a combination
of analytic expectation and experience.  First we define
\be l_\Lambda = 40\sqrt{ {|\Omega_\Lambda-\Omega_\Lambda^*| \over
\Omega_m}},
\ee
\be l_{\rm curv} = {\rm
min}\left[200\left({\Omega_K\over\Omega_m}\right)^{1/3},120\right], \ee
with $l_{\rm curv}$ defined for $\Omega_{\rm K}\geq 0$, and then we set
$l_{\rm late}={\rm max}(l_{\rm curv},l_\Lambda)$. For closed models we
set
\be
l_{\rm late} = {\rm max}\left[ 25 (-\Omega_K)^{1/3}, l_\Lambda
\left(1-4\frac{\Omega_{\rm K}}{\Omega_m}\right)^{-1}\right],
\ee
which takes into account the effect of both $\Lambda$ and
curvature.\footnote{
For closed models in the second--grid approach we
we simply set $l_{\rm late}=\rm{max}(40,l_\Lambda)$ which works well
for the parameter range under consideration
($|\Omega_{\rm K}| < 0.3$).}.

Our reasoning is that below $l_{\rm late}$, ISW and curvature effects
become important.
In principal $l_s \equiv \pi/\theta_s$ is another important scale,
above which acoustic modifications to the intrinsic temperature on the
last--scattering surface become important.  We avoid extending the late
grid to $l > l_s$ because this allows us to grid coarsely in $\w_b$.
Fortunately $l_{\rm late}$ is always less than $l_s$.  We choose to make
the switch at $l_{\rm late}$ instead of $l_s$ so that he low--$\ell$
grid can be coarse in $\w_m$; cutting at higher $\ell$ would require
finer grids in $\w_m$ to accurately describe the early ISW effect.

To calculate $\Delta_l(k)$ for the target model from our
$n$--dimensional grid we first locate the $2^n$ grid points of the
surrounding hypercube.  Then for each of these $2^n$ $\Delta_l(k)$
vectors we spline--interpolate (and quadratically extrapolate where
necessary) on to the $k$ values of the target model.  This step is
necessary because the $k$ values of the grid differ from grid point to
grid point.  A uniform set of $k$ values is not desirable since
different models have different $k$--spacing requirements for fixed
accuracy specification.  For closed models a uniform set of $k$ values
is impossible due to the discrete nature of the spectrum.

The grid is inefficient (in computing time and storage resources)
if a lot of the grid points are for models which are far from
observationally viable.  Since we use rectangular grids, this
means we would like to choose parameters such that their viable
ranges are independent of the values of the other grid parameters.
A systematic way to do this would be to use the eigenvectors of
the parameter Fisher matrix for some particular experiment
\citep{efstathiou99}.

We have not pursued this grid efficiency systematically, but
rather have made the physically motivated choice of our low--$\ell$ grid
parameters as $\w_b$, $\ln{\w_m}$, $\sqrt{\Omega_\Lambda/\Omega_m}$ and
$\w_K$. Of course, even with this parameterization, we are still
free to find the parameter eigenvectors. Perhaps doing so will further
increase the efficiency of the grid and we may incorporate this in
future DASh implementations.  Not only will eigenvectors provide
the advantage of a grid with a rectangular region of viable
models, but there is also an advantage in having the
well--determined parameter combinations decoupled from the
poorly--determined combinations.  Variations in the
well--determined ones (over the range of their compatibility with
data) are most likely to reproduce a highly linear response in
$\C_l$ and thus would require very few grid points (only two if the
response were exactly linear).  Variations in the
poorly--determined parameters will generate a {\it non}--linear
response, but we will not need to model these responses as accurately
so once again will not need many grid points.

\section{Accuracy}

To characterize the accuracy of DASh we have compared a suite of
thousands of models calculated using CMBfast and compared them
with the same models as calculated by DASh.  The CMBfast
calculations were done with very dense $k$--spacing, effectively
eliminating $k$--spacing as a source of error.  The DASh
calculation used the grids as described in Table 1.  We created
our suite of comparison models by first considering 
all possible models with parameter values $\w_b =
0.0145,.018,.022,.028$, $\w_d = .06,.09,.15,.2$, $\Omega_\Lambda =
0,0.2,0.4,0.5,0.6,0.7,0.8$, \\$\Omega_K =0,\pm 0.01,\pm
0.05,\pm0.13, \pm0.21,\pm0.27$, $z_{\rm ri} = 0,5,10$ and $n =
0.95,1,1.05$ where $n$ is the scalar power--spectrum power--law index.
Of these models, those within the grid boundaries of
Table 1 and having $0.3 < h < 1$ and a first acoustic peak with $160 <
l < 280$ were placed in the comparison suite.  The suite
contains 313 models with $z_{\rm ri} = 0$ and $\Omega_K=0$, 937
models with $\Omega_K=0$ and 6,823 models total. We refer to the
differences between the two calculating tools simply as
`differences' rather than `DASh errors' because some of the
differences are due to errors in CMBfast.  Also some errors result
in no difference because they are common to DASh and CMBfast.  We
have not attempted the more challenging task of providing an
absolute measure of accuracy.
 \begin{figure}[htbp]
  \begin{center}
    \plotone{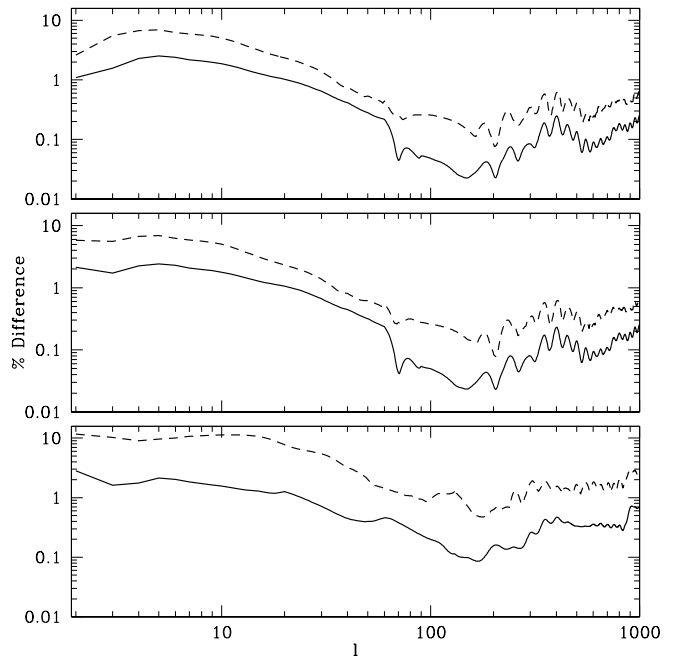}
    \caption{Differences between sDASh and CMBfast.
    Solid (dashed) lines show rms (maximum) percentage
    differences. The bottom panel is for all 6,823 models in the
    comparison suite (see text).  The middle panel is for the 937
    flat models in the suite.  And the top panel is for the 313
    flat models with $z_{\rm ri}=0$.    }
    \label{fig:diff-s}
  \end{center}
\end{figure}
 \begin{figure}[htbp]
  \begin{center}
    \plotone{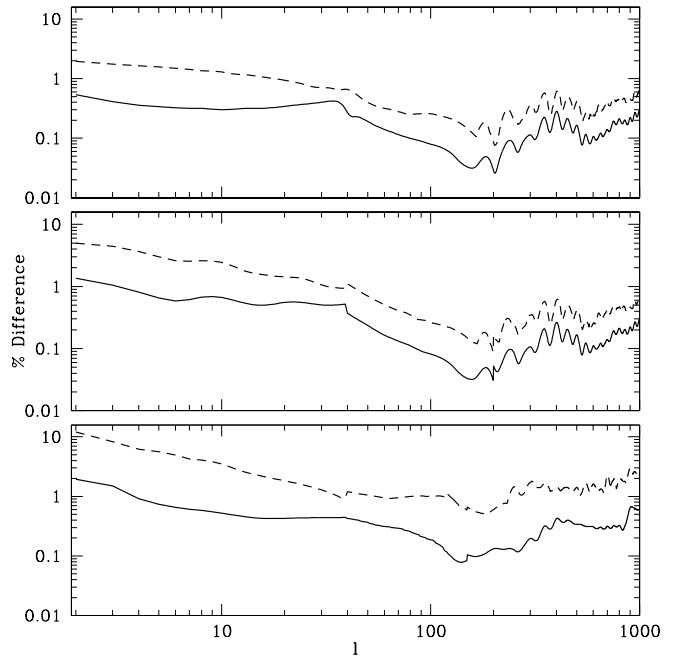}
    \caption{Differences between gDASh and CMBfast. See caption
    for Figure 1.}
    \label{fig:diff-g}
  \end{center}
\end{figure}

The results are shown in Figure 2 for sDASh and
Figure 3 for gDASh.  In both figures we have
shown the maximum and average differences for flat models (top
panels), for flat models with reionization (middle panels), and for
flat and curved models with reionization (bottom panels).  One can see
that in every case the rms differences are below 1\% for $l \ga 30$
for sDASh and $l \ga 10$ for gDASh.
The rms fractional difference (for $\ell < 1000$) is in all cases 
below the cosmic variance error of $\sqrt{1/l}$. 

The size of the differences relative to cosmic variance
is encouraging, but unlike cosmic variance the
calculational errors that give rise to these differences may be 
correlated from $\ell$ to $\ell$.  Therefore the target level is
closer to $1/l$ than $\sqrt{1/l}$ (see \citet{knox98} and
\citet{efstathiou99}) since $1/l$ is the cosmic variance error on
a band of $C_l$ of width $l$.  At all $\ell$ values, DASh 
either meets this more conservative criterion or has sub--percent
level differences with CMBfast.  

DASh errors are certainly small enough to be insignificant for
parameter estimation from current data.  But we do not yet understand
the impact of sub percent level errors on parameter--determination
from all--sky surveys such as the Microwave Anisotropy Probe ({\it
MAP})\footnote{http://map.gsfc.nasa.gov}.  This is under investigation.
The interpretation of the sub--percent level differences with
CMBfast is complicated by the fact that the CMBfast calculation
for some models can have percent level errors.

There are also errors due to neglect of non--linear effects. The
most important of these are due to gravitational lensing and
peculiar velocities of the reionized inter--galactic medium.  We
discuss including these effects in the Extensions section.

\subsection{High $\ell$ errors}

At high $\ell$ the top two contributors to error in the DASh
calculations are interpolation error (at a level of about 0.5\% by
grid design) and error from the projection approximation.
The chief cause of differences in the curved case
is due to error in our CMBfast calculations arising from
inaccuracies in the hyperspherical Bessel functions.

\begin{table*}[hbt]\small
\caption{\label{table:grid}}
\begin{center}
{\sc Grid Properties}\\
\begin{tabular}{ccccccccccc}
\tableskip\hline\hline\tableskip
 & $\sqrt{\Omega_\Lambda/\Omega_m}$ & $\w_K$  & $\ln{w_m}$  & $w_b$  &
$\tau$ &total& $k_{\rm max}\eta_0$ & $n_k$ & $n_l$ & Storage (Mb)\\
\tableskip\hline \tableskip
Low-$l$ $\Delta_l^2(k)$  & 8 &32 & 8 &2 & 1 & 2,330& 1600 & 1600 & 22 & 570\\
High-$l$ $\Delta_l^2(k)$ & 1 &1 & 32 &8 & 1 &256 & 6000 & 4500 & 60 & 250\\
$F_l$ (reionization) & 1 &1 & 4 &4 & 8 &128 & 1600 & 1500 & 24 & 0.1\\
\tableskip \hline \tableskip
range & 0:2.44 & -0.15:0.16 & $\ln(0.06):\ln(0.28)$ & 0.01:0.03 & 0:0.3\\
\tableskip\hline
\end{tabular}\\[12pt]
\begin{minipage}{5.2in}
NOTES.---%
Entries in the middle rows to the left of the `total' column show
the number of values given for each parameter in the various
grids.  The `total' column shows the total number of models
in each grid.  The low $\ell$ total is less than the product of the
other entries in the row because models that do not satisfy $0.1 < h < 1.1$
and $\Omega_m > 0$ and $0\leq \Omega_\Lambda \leq 0.9$ are not calculated.
All grids were given the same parameter ranges, shown
in the bottom row.  The High--$\ell$ $\Delta_l^2(k)$ and $F_l$ grids
were evaluated at fixed $\Omega_\Lambda = 0.6$ and 0 (respectively),
$w_K = 0$ and $\tau=0$.  The last four columns show the maximum
$k$ values, number of $k$ values, number of $\ell$ values calculated
and the storage requirements (in Megabytes) for each grid.
\end{minipage}
\end{center}
\end{table*}

\begin{figure}[htbp]
  \begin{center}
    \plotone{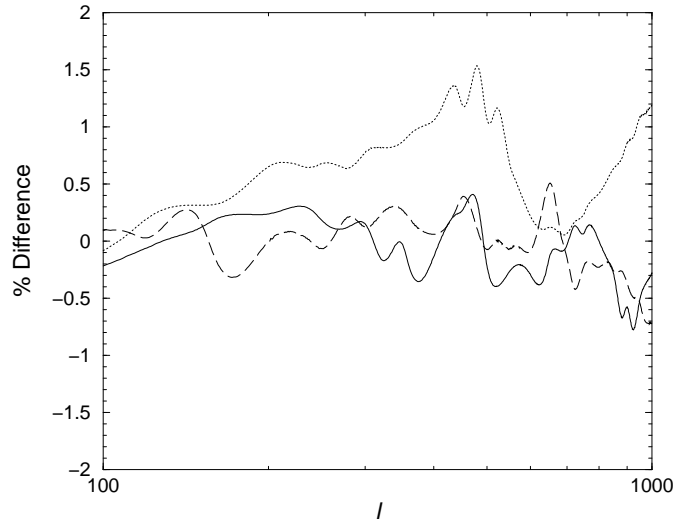}
    \caption{Projection errors for different components of the
anisotropy. The target model is flat with $\Omega_\Lambda=0.9$,
$\w_b=0.02$ and $\w_m=0.12$, while the fiducial model is flat and
has $\Omega_\Lambda=0$. The ordinate shows the relative error (in
\%) for different components of the anisotropy produced at or
close to the last scattering surface. The solid line shows the
projection error for the total anisotropy, the dashed line for
that of the Doppler contribution while the dotted line shows the
error in projecting the ISW effect to the target model angular
diameter distance.}
    \label{fig:shifting}
  \end{center}
\end{figure}

The projection approximation works remarkably well and we now turn
to understanding that success. This can be seen from Figure
\ref{fig:shifting} where we have plotted the differences between
two flat models, shifted to correct for their 25\% difference 
in angular diameter distance to the last scattering
surface. Both the fiducial and target models in Figure
\ref{fig:shifting} are flat and were calculated using CMBfast.
Even for this large a shift, the errors at high $l$ are $\la
0.5\%$. We will argue that these differences must be due
to numerical errors other than the projection approximation.
We expect the projection to work just as well
for curved models.

The scaling of $l(l+1)C_l$ with angular--diameter distance can be
derived if one assumes the emission is from isotropic sources on
an infinitesimally thin last--scattering surface. Then the
correlation of temperatures at a given angular separation is equal
to the correlation of temperatures on the last--scattering surface
with a given physical separation. Therefore one can determine the
angular correlation function of one model, from the angular
correlation function of another, as long as both models have the
same physical conditions before and at last--scattering.
Specifically, spatial perturbations to the effective photon
temperature monopole, $\Theta_0+\Psi$, with angular correlation
function $C(\theta)$ at angular diameter distance $r$, have
correlation function $C'(\theta) = C(\theta r'/r)$ at angular
diameter distance $r'$.  The effect on the power spectrum is to
have ${\cal C}'_l \equiv l(l+1)C'_l/(2\pi)={\cal C}_{l'}$ where
$l' = r/r'l$ since: \bea
{\cal C'}_l &=& l(l+1) \int d(cos\theta) C(\theta r'/r) P_l(cos\theta) \nonumber\\
& \simeq & l^2 \int \theta d\theta C(\theta r'/r) J_0(l\theta) \nonumber \\
& = & l'^2 \int dx x C(x) J_0(l'x) = {\cal C}_{l'}
\eea
where the approximation is accurate for  $\theta << 1$.

We do not expect the thickness of the last--scattering surface to be a
significant source of error.  We find analytically that for thickness
$t$ and angular diameter distance $d$ shifted to $d+\delta d$, the
error in $C_l$ is of the order $(\delta d/d)(t/d)^2l^2 d^2 C_l/dl^2$
which is only important at the sub-$0.1\%$ level.  

Anisotropies are generated soon after last scattering by the early
ISW effect due to the decay of the gravitational potential in the
presence of radiation.  Thus for the early ISW effect there is a much
thicker ``last--scattering surface'' and the projection approximation
does not work as well as it does for the other early sources.  However,
the region in $\ell$ space where the approximation is worst is also
where it is a highly subdominant contribution to the total anisotropy.

Note that velocity perturbations on the last--scattering surface
are a significant contribution to the anisotropy, and are not
isotropic sources on the last--scattering surface.  
Velocity correlations in
three--dimensions decompose into the correlation between
components perpendicular to their separation vector and the
components parallel to their separation vector.  At small scales,
the radial direction is nearly perpendicular to the separation vector
and since the radial component is all that is important for the 
Doppler effect, we are primarily sensitive to the perpendicular 
component of the velocity correlation.  This component projects
like the monopole and the result is that dipole sources do
not introduce much error, as can be seen in Fig.~\ref{fig:shifting}.  
At larger scales the
parallel components of the velocity correlation also become important, 
but the velocity contribution is sufficiently small at larger 
angular scales that the resulting shifting errors are negligible.

A comparison of the open models to flat fiducial models using
CMBfast shows that the high $l$ shifting differences are much
larger than the errors we expect analytically. The 1\% to 2\%
differences are due to errors in the CMBfast calculation,
presumably in the hyperspherical Bessel functions it uses. We
compared the CMBfast outputs (at fixed $\w_m$ and $\w_b$) for a
flat model, a model with $\Omega_{\rm K}=-0.002$ and another with
$\Omega_{\rm K}=0.001$. (All 3 models had $\Omega_\Lambda=0$.) The
differences in the angular diameter distances are less than
$0.1\%$ and hence we would expect very little difference in $C_l$.
Instead we find 1--2\% changes to $C_l$ at high $\ell$ between the 
flat and curved models, presumably due to difficulty in calculating the
high $l$ hyperspherical Bessel function accurately. Hence we conclude
that  projecting from a flat fiducial model to a curved target model is
more accurate than the direct calculation using hyperspherical Bessel
functions.

Although the current difference plots show we may not have met the
more conservative criterion of having fractional errors below $1/l$,
we are optimistic that we can get there.  As data improve DASh will be 
able to improve along with them.  Better data mean smaller viable regions 
of parameter space, so the grid boundaries can shrink, allowing for 
greatly decreased interpolation error without increasing the 
number of models in the grid.  Projection errors, already small,
will be further reduced since the shifts from the fiducial model
will be smaller.  

\subsection{Low $\ell$ errors}

The gDASh can be made
arbitrarily accurate at low $\ell$ 
by decreasing the parameter grid spacing. This is
not the case for sDASh where the accuracy is
limited by the approximations made.  
Of course, even in the second grid
method, decreasing the parameter grid spacings requires more
computational resources, primarily in being able to pre--compute
and store large number of $\Delta_l(k)$ files. 

The default grid--spacings in DASh have been chosen so that the
interpolation errors are $\la 0.5\%$ at high-$\ell$. For low $\ell$,
where percent accuracy is not important, the constraints imposed on the
grid-spacings are more lenient. All the low $\ell$ sDASh errors
are due to interpolation.

The sDASh errors at $\ell < 10$ are larger than the
gDASh errors. However, the conservative $1/l$ cosmic variance
criterion is much easier to meet here.  Maximum
errors of the order of 10\% can be tolerated. The maximum errors always
come from the largest values of $|\Omega_{\rm K}|$.  A large part of the
error is due to our approximate calculation of the cross-term between the
early contributions and the late ISW contribution.  This 
could be improved as discussed earlier.

Another source of error in the semi-analytic calculations at very low
$l$ ($l \la 5$) is the inability to carry out the projection as outlined
in Eq.~\ref{eqn:projectlarge}. Let $r'$ and $r$ be the coordinate
distances to the last scattering surface of the target and fiducial
models respectively. If $r'>r$, then there is no angle in the fiducial
model that projects on to $180^o$ in the target model. Hence one cannot
obtain $C_l$ by a Legendre transform. Choosing our fiducial model to be
$\Omega_\Lambda=0.6$ exacerbates this problem. However that choice for
the fiducial $\Omega_\Lambda$ is justified by the aim of getting the
high-$l$ spectra very accurately for models close to the fiducial
one. Note that choosing a non-zero $\Omega_\Lambda$ to be the fiducial
model introduces another source of error into the low $l$ semi-analytic
calculations since once needs to subtract off the ISW contribution due
to the fiducial $\Omega_\Lambda$. The error introduced due to this is
about the same as the error in calculating the flat models.
One solution to these problems is to implement another high-$l$ grid with
$\Omega_\Lambda^*=0$. This increases the pre-computing time and the
storage requirements. We have not implemented this second
grid but instead we approximate the projection effect and tolerate the
error in the subtraction of the $\Omega_\Lambda^*$ ISW effect. For the
projection, we use $l'(l'+1)=l(l+1)*r'^2/r^2$ to get to $l'=4$ and then
use the fact that at these large angular scales the spectrum is mostly
sourced by the SW effect. Hence the lower multipoles can be obtained
from $l'=4$ by assuming that the ratio of ${\cal C}_l/{\cal C}_4$ is
just given by the corresponding ratio of the integrals over the
hyperspherical Bessel function squared times the power spectrum. 
We pre--compute these integrals assuming a scale--invariant power spectrum;
deviations from scale--invariance are unlikely to be a significant
source of error.

\subsection{Speed}

We used a personal computer with Pentium IV processors and version 2.96
of the GNU gcc g77 compiler for all our calculations.  Computation of
the high--$\ell$ $\Delta_l(k)$ grid took $\sim 4$ hours, of the low--$\ell$
$\Delta_l(k)$ grid took $\sim 40$ hours and of the $F_l$ grid 
took $\sim 2$ hours.
After this pre--computing, gDASh requires 1.5 seconds and sDASh
requires 2 to 3 seconds to calculate $C_l$.  
For comparison, CMBfast (with the same $k_{\rm max}$ and
$n_k$ as used to precompute our grid---see Table 1) takes 50 seconds
for flat models and about 90 seconds for $\Omega_K \ne 0$ models.

Significant time is spent reading the grid files each time a new
model is computed.  The grids we used (Table 1) could be stored in less
than a Gbyte of RAM, and doing so may speed DASh up by a factor of two.

\section{Polarization and Tensor Spectra}

The polarization power spectrum is calculated in DASh in a manner
similar to the temperature spectrum. We store the polarization analogs
of $\Delta_l(k)$ in the high--$\ell$ grid and use the same formulae for
angular--diameter distance shifting and re--ionization
suppression. The grid sizes for the polarization $\Delta_l(k)$ are
smaller by about a factor of 2 because polarization $\Delta_l(k)$ is
non-negligible over a smaller range of $k$ for any given $\ell$. The
number of grid points required to achieve percent level accuracy is
comparable to the temperature grid. 

Reionization results in a new peak in the power spectrum at very low
$\ell$ \citep{zaldarriaga97}. It is the only important late--time
effect; there is no analogous ISW effect for polarization. We have found
a simple scaling relation that allows it to be calculated rapidly: 
\begin{eqnarray}
\label{eqn:pol}
{\cal C}_{El'} & = & {\cal C}_{E_* l}
\frac{(1-e^{-\tau})^2}{(1-e^{-\tau_*})^2}
\left(\frac{\tau_*}{\tau}\right)^{(0.2-\tau_*/3)}
\left(\frac{A}{A_*}\right)
\left(\frac{2}{l_{\rm pivot}}\right)^{n-1}, \nonumber \\ 
{\cal C}_{Cl'} & = & {\cal C}_{C_* l}
\frac{(1-e^{-\tau})}{(1-e^{-\tau_*})}
\left(\frac{\tau_*}{\tau}\right)^{0.2}
\left(\frac{A}{A_*}\right)
\left(\frac{2}{l_{\rm pivot}}\right)^{n-1},
\end{eqnarray}
where ${\cal C}_{E l}$ is auto-correlation of the E (electric or scalar)
mode of polarization and ${\cal C}_{C l}$ is the (E mode) polarization --
temperature cross-correlation. 
The initial potential power spectrum is taken to be 
$P(k) = A\ (k/k_{\rm pivot})^{n-4}$, and 
$l_{\rm pivot} \equiv (6000\ {\rm Mpc}\ k_{\rm pivot})
/ (\omega_m (1+z_{\rm ri}))^{1/2}$.
The angular diameter distance shifting is achieved through the relation
$l' = l ({l'}_r/l_r)$ for ${\cal C}_{El}$ and 
$l' = l ({l'}_r+0.5)/(l_r+0.5)$ for ${\cal C}_{Cl}$.
The fiducial model, denoted by ``$*$'', which we have used in our
scaling relations has $\tau=0.15$, $\omega_b=0.02$, $\omega_m=0.16$ and
$\Omega_\Lambda=0.65$.   
Since most of the optical depth to reionization is generated prior
to the onset of curvature or dark energy domination there is
no dependence on $\Omega_k$ or $\Omega_\Lambda$.  Also, since the
relevant wavelengths are much larger than the horizon at last scattering
there is no dependence on $\omega_b$.  Eq.~\ref{eqn:pol} is accurate
to $\sim 10\%$ near the peak of the reionization ``bump''.
When data warrant higher accuracy, a low $\ell$ grid can be created 
just as in the temperature case.

For the tensor temperature power spectrum, perturbations 
are continually generating temperature anisotropy along the
line--of--sight so there is no useful late--time/early--time
split and the simple angular--diameter distance projection
corrections do not apply.  
However, a sufficiently accurate ($~3\%$) {\em fit} to the tensor
temperature power spectrum  already exists  \citep{turner96} and we have
included this in DASh. 
Note that polarization generation from tensor perturbations only occurs at
last--scattering and after reionization so these could be
included, when data warrant, in a manner analogous to the scalar case.

\section{Extensions}

There are a number of ways in which DASh could be extended.
Additional effects can be included such as lensing, the
Ostriker--Vishniac effect, patchy reionization, gravitational
waves and dark energy with $p/\rho \ne -1$.  DASh could
also be extended to calculate the CMB
polarization power spectra\footnote{This has in fact already
been done for a preliminary version of DASh (B. Gold, private
communication)}.

Lensing of CMB photons by mass inhomogeneities on our past light
cone leads to a smoothing of the power spectrum by a
smoothing kernel $W^l_{l'}$ such that $C_l = \sum_{l'}W^l_{l'}C_{l'}$.
This correction can be calculated rapidly and accurately as demonstrated
by \citet{zaldarriaga98} who calculate it in real space.  This
smoothing by lensing is a very significant effect, leading to
corrections at the several percent level at $l = 1000$ and tens of
percent corrections at $l=3000$ \citep{metcalf98}.  These corrections
are useful because they break the curvature, $\Omega_\Lambda$
degeneracy \citep{metcalf98}.

Above we have described how reionization generates new anisotropy on
very large scales as photons pick up some of the peculiar momentum of
the electrons via scattering.  Although the linear theory contribution
from this effect is very small at small angular scales, the second
order contribution, called the Ostriker--Vishniac (OV) effect
\citet{ostriker86}, can be the dominant source of anisotropy at $\ell
\ga 3000$.  The OV contribution is ${\cal C}_l \simeq 5 \muK^2$
according to a numerical calculation by \citet{springel01} and
an analytic calculation by \citet{ma01}; also see
recent forecasts for secondary anisotropy by \citet{aghanim02}.  Since
it is primarily a second--order effect it is especially sensitive to
the amplitude of the fluctuations and is therefore not sensitive to
events at high redshift.  In particular, since $z_{\rm ri} > 6$ it is
insensitive to $z_{\rm ri}$.  Semi--analytic means of rapidly
calculating the power spectrum from the OV effect exist and could
easily be included in DASh \citep{hu96,jaffe98}.

The transition from neutral IGM to ionized IGM is likely to go through
a ``patchy'' period in which the two phases are spatially mixed.  If
patches of reionized IGM are small enough (comoving linear extent less
than about 20 Mpc) then the first order contribution to anisotropy
from scattering off of electrons does not experience cancellations and
can be large \citep{aghanim96,gruzinov98,knox98}.  This contribution
from patchy reionization is proportional to the redshift width of the
transition and $(1+z_{\rm ri})^{3/2}$ \citep{gruzinov98}.  
A small $z_{\rm ri}$ means
that patchy re--ionization is almost certain to be subdominant
compared to the non--patchy contribution (although see
\citet{aghanim02} who assume a large typical patch size and find the
patchy phase may contribute significantly).



The current implementation of DASh assumes the dark energy is
a cosmological constant (i.e., $p = -\rho$).  Whether the dark
energy is a cosmological constant or not is a matter of profound
importance in cosmology and fundamental physics.  Dark energy models
based on slowly--rolling scalar fields have $w \equiv p/\rho \ne -1$
and can in principle be observationally distinguished from a cosmological
constant.  The effect of dark energy on the CMB is solely gravitational;
all we have to do is calculate how the dark energy affects $\Psi$.
\citet{ma99} have shown that the effect of dark energy on
the matter evolution can be factorized into one function of $k$ and
one function of time.  We can use their fitting formulae for these
functions to rapidly calculate $\Psi$ at late times and therefore the
ISW contribution to anisotropy.

\section{Conclusions}

We have described a fast and accurate method for calculating
angular power spectra from the parameters of adiabatic models. Our
implementation, DASh, is publicly available.  Extensions for
polarization, tensor modes and lensing are straightforward.

The speed of DASh will be useful for parameter estimation from CMB
power spectrum data which typically requires more than hundreds of
thousands of models to be calculated.  DASh is particularly
advantageous compared to grid--based $C_l$ calculations for
parameter estimation techniques which require calculation of
$C_l$ at random points in the parameter space, such as the
MCMC method used in \citet{knox01} and described in \citet{christensen01}.
A preliminary version of DASh, valid only for flat models, was
used in \citet{knox01}.  Since we pre--compute
a transfer function instead of $C_l$ DASh's advantage will
be greatest for applications with large numbers of primordial
power spectrum parameters---such as attempts to reconstruct
this spectrum from data \citep{wang00}.

\acknowledgements
We are grateful to U. Seljak and M. Zaldarriaga for making CMBfast
publicly available, S. Colombi, M. Tegmark and M. Zaldarriaga for useful
conversations, and A. Kosowsky for his hyperspherical
Bessel function routines.  We used CAMB \citep{lewis00} many times as we
worked through various accuracy tests.  LK thanks the IAP for
their hospitality during early stages of this work.  This work was
supported by NASA grant NAG5-11098.


\end{document}